\pdfoutput=1
\documentclass[a4paper,american,floatfix,pdftex,superscriptaddress,twocolumn,twoside,%
aps,prl,%
citeautoscript,
longbibliography,%
reprint,
]{revtex4-1}%
\usepackage{amsfonts,amsmath,amssymb}
\usepackage{commath}
\usepackage[T1]{fontenc}
\usepackage{graphicx}%
\usepackage[utf8]{inputenc}
\usepackage{microtype}
\usepackage[obeyFinal,textsize=footnotesize]{todonotes}
\usepackage{xspace}
\usepackage{hyperref, hypernat}
\usepackage[todos]{CMImacros}

\newcommand{\cfeldesy}{\affiliation{Center for Free-Electron Laser Science, Deutsches
      Elektronen-Synchrotron DESY, Notkestraße 85, 22607 Hamburg, Germany}}%
\newcommand{\uhhcui}{\affiliation{Center for Ultrafast Imaging, Universität Hamburg, Luruper
      Chaussee 149, 22761 Hamburg, Germany}}%
\newcommand{\uhhphys}{\affiliation{Department of Physics, Universität Hamburg, Luruper Chaussee 149,
      22761 Hamburg, Germany}}%
\newcommand{\ucl}{\affiliation{Department of Physics and Astronomy, University College London, Gower
      Street, WC1E 6BT London, United Kingdom}}%
\newcommand{\jkemail}{\email[]{jochen.kuepper@cfel.de}}%
\newcommand{\ayemail}{\email[]{andrey.yachmenev@cfel.de}}%
\newcommand{\cmiweb}{\homepage{https://www.controlled-molecule-imaging.org}}%

\DeclareMathOperator{\atantwo}{atan2}
\DeclareMathOperator{\sign}{sign}

\newcommand*{\ee}{\ensuremath{\text{\emph{ee}}}\xspace}%

\begin{document}
\title{Field-induced diastereomers for chiral separation}%
\author{Andrey Yachmenev}\ayemail\cfeldesy\uhhcui%
\author{Jolijn Onvlee}\cfeldesy%
\author{Emil Zak}\cfeldesy%
\author{Alec Owens}\cfeldesy\uhhcui\ucl%
\author{Jochen Küpper}\jkemail\cmiweb\cfeldesy\uhhcui\uhhphys%
\date{\today}
\begin{abstract}\noindent%
   A novel approach for the state-specific enantiomeric enrichment and the spatial separation of
   enantiomers is presented. Our scheme utilizes techniques from strong-field laser physics,
   specifically an optical centrifuge in conjunction with a static electric field, to create a
   chiral field with defined handedness. Molecular enantiomers experience unique rotational
   excitation dynamics and this can be exploited to spatially separate the enantiomers using
   electrostatic deflection. Notably, the rotational-state-specific enantiomeric enhancement and its
   handedness is fully controllable. To explain these effects, we introduce the conceptual framework
   of \textit{field-induced diastereomers} of a chiral molecule and perform robust quantum
   mechanical simulations on the prototypical chiral molecule propylene oxide (C$_3$H$_6$O), for
   which ensembles with an enantiomeric excess of up to $30~\%$ were obtained.
\end{abstract}
\maketitle

\noindent%
Chirality is central to many chemical and biological processes. Its significance is emphasized by
the fact that life on Earth is based on chiral biomolecules, which are all naturally selected with a
single handedness that determines their functionality. Chiral molecules occur in structural forms
known as enantiomers, which are mirror images of one another and, therefore, non-superimposable by
translation and rotation. Given that molecular enantiomers have identical physical properties,
neglecting the so-far unobserved effects of parity-violating weak interactions, but often strikingly
different chemical behaviour, methods to distinguish and/or separate enantiomers are extremely
important, particularly in areas such as drug design and pharmacology. Sources of cold chiral
molecules in distinct enantiomeric states could improve measurements of electroweak
interactions~\cite{Daussy:PRL83:1554, Quack:ARPC59:741, Schnell:FD150:33} and studies of collisional
dynamics with chiral molecules~\cite{Lombardi:JPCM30:063003}.

Recently, a number of robust techniques for separating and purifying enantiomers in the gas phase
from mixed chiral samples have been proposed~\cite{Eibenberger:PhysRevLett118:123002,
   Perez:ACIE56:12512, Banerjee-Ghosh:Science360:1331, Wang:NatComm5:3307, Hayat:PNAS112:13190}.
These developments are a consequence of the increased precision and control offered by gas-phase
chemistry, which is providing exciting advances in the analysis and manipulation of chiral
molecules. Sophisticated methods are now available for establishing the absolute configuration and
enantiomeric excess (\ee), for example, employing phase-sensitive microwave
spectroscopy~\cite{Patterson:Nature497:475, Domingos:ARPC69:499}, photoelectron circular
dichroism~\cite{Janssen:PCCP16:856}, Coulomb explosion imaging~\cite{Pitzer:Science341:1096,
   Herwig:Science342:1084}, high-harmonic generation~\cite{Cireasa:NatPhys11:654}, or
attosecond-time-resolved photoionization~\cite{Beaulieu:Science358:1288}.

Here, we present a novel approach for coherent enantiomer-specific enrichment of rotational state
populations of chiral molecules coupled to state-specific electric-field manipulation. Our scheme
utilizes an optical centrifuge~\cite{Karczmarek:PRL82:3420}, which is a linearly polarized light
pulse that performs accelerated rotation about the direction of propagation, in conjunction with a
static electric field along the light propagation direction. This yields a chiral electric field
that induces unique rotational excitation dynamics in the different enantiomers . The
rotational-state-specific enantiomeric enhancement for either the left- or right-handed enantiomer
is fully controllable by changing the duration of the optical centrifuge pulse, and it is
subsequently transformed into spatial enhancement using the electrostatic
deflector~\cite{Chang:IRPC34:557}.

\begin{figure}
   \includegraphics[width=\linewidth]{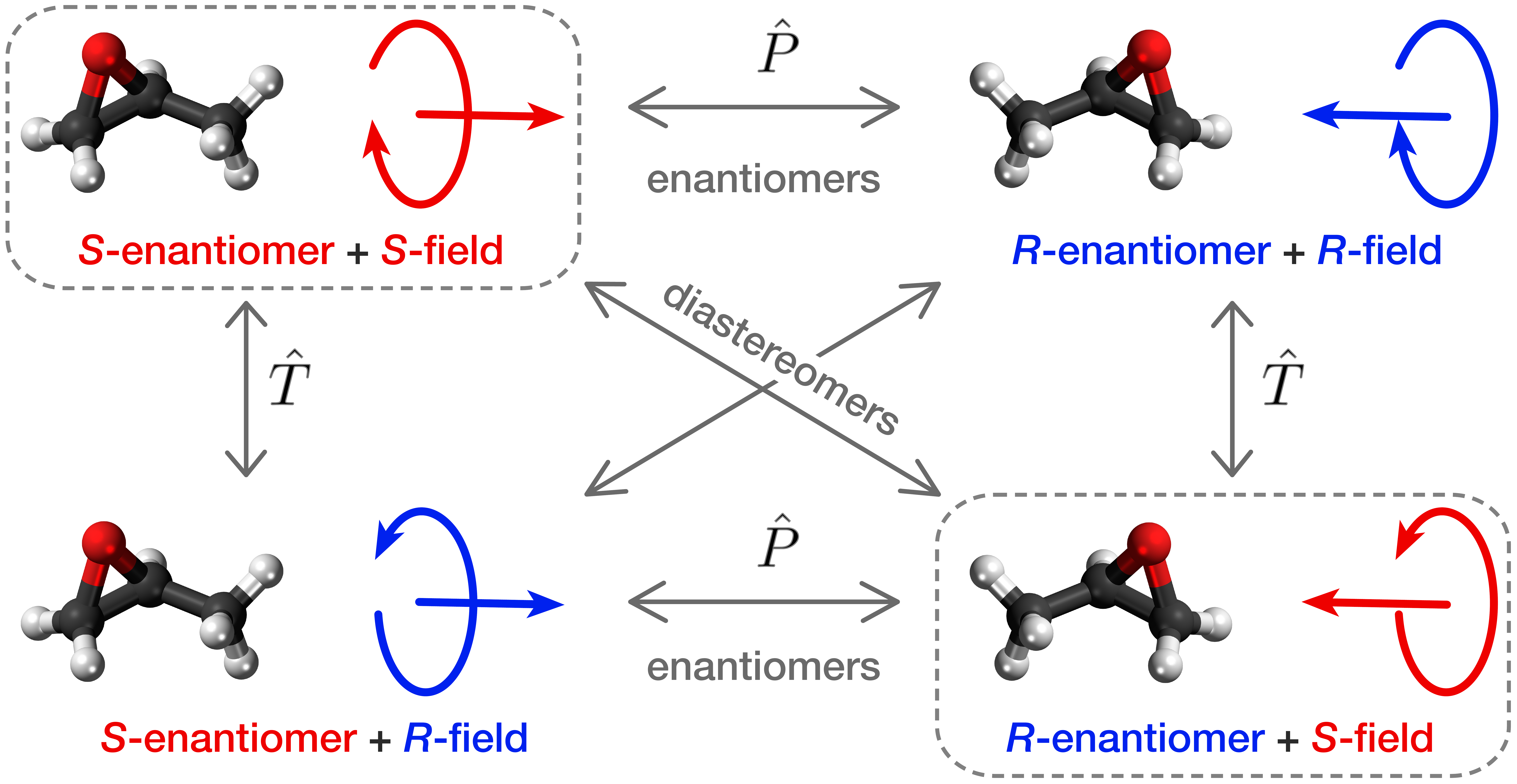}
   \caption{Field-induced diastereomers of a chiral molecule interacting with a chiral electric
      field, \ie, a rotating optical field in conjunction with a static electric field along the
      rotation axis of the optical field.}
   \label{fig:concept}
\end{figure}
Similar to classical methods of chiral resolution, where a pure enantiomer of another chiral agent
is introduced to form diastereomeric complexes with distinct physiochemical properties, we can
create \textit{field-induced diastereomers} of a chiral molecule, depicted in \autoref{fig:concept},
by placing its enantiomers in a chiral electric field with defined handedness. The symmetry
properties of the field-induced diastereomers are such that both, the molecule and the field
enantiomers, are interconverted by the parity inversion operator $\hat{P}$, but time inversion
$\hat{T}$ interchanges only the field enantiomers. Hence, two chiral centers emerge in the presence
of the electric field with defined handedness, shown in the dashed panels in \autoref{fig:concept},
and as for classical chemically-bound diastereomers, with distinct energies, the rotational dynamics
of the enantiomers will differ.

The interaction of the enantiomers with a chiral electric field is described by the following
time-dependent potential,
\begin{align}
  \label{eq:v_ocdc}
  V(t) &= \frac{1}{4}\epsilon(t)^2\cos^2(\omega t)V^\text{oc} + \frac{1}{\sqrt{2}}\epsilon_\text{dc}V^\text{dc},
\end{align}
with the optical-centrifuge contribution
\begin{align}
  \label{eq:v_oc}
  V^\text{oc} &= -2\alpha D_{0,0}^{(0)*} \\ \nonumber
              &- \Delta{\alpha} \left(e^{2i\beta t}D_{-2,0}^{(2)*} + e^{-2i\beta t}D_{2,0}^{(2)*}- \sqrt{\frac{2}{3}}D_{0,0}^{(2)*}\right) \\ \nonumber
              &+ \alpha_{xz}\left(  e^{2i\beta t}\mathcal{A}_{-2,1}^{(2)} +  e^{-2i\beta t}\mathcal{A}_{2,1}^{(2)} -\sqrt{\frac{2}{3}}\mathcal{A}_{0,1}^{(2)}  \right) \\ \nonumber
              &+ i\alpha_{yz}\left(  e^{2i\beta t}\mathcal{S}_{-2,1}^{(2)} +  e^{-2i\beta t}\mathcal{S}_{2,1}^{(2)} -\sqrt{\frac{2}{3}}\mathcal{S}_{0,1}^{(2)}  \right)
\end{align}
and the dc-field component
\begin{equation}\label{eq:v_dc}
   V^\text{dc} = \mu_{x}\mathcal{A}_{0,1}^{(1)} + i\mu_{y}\mathcal{S}_{0,1}^{(1)} - \sqrt{2}\mu_z D_{0,0}^{(1)*}.
\end{equation}
Here, $\epsilon(t)\cos(\omega{}t)$ describes the linearly polarized carrier ac field of the optical
centrifuge with the pulse envelope function $\epsilon(t)$ and angular frequency $\omega$ and the
acceleration of the centrifuge circular rotation $\beta$. $\epsilon_\text{dc}$ is the dc field
strength. The electric field tensors $\mu_{\gamma}$ ($\gamma=x,y,z$),
$\alpha=(\alpha_{xx}+\alpha_{yy}+\alpha_{zz})/3$ and
$\Delta{\alpha}=(2\alpha_{zz}-\alpha_{xx}-\alpha_{yy})/\sqrt{6}$ are the permanent dipole moment,
average static polarizability, and the static polarizability anisotropy, respectively, in the
principal axis of inertia molecular frame. The symbol $D_{m,k}^{(J)*}$ denotes the
complex-conjugated Wigner $D$-matrix, $\mathcal{A}_{m,k}^{(J)}= D_{m,k}^{(J)*} - D_{m,-k}^{(J)*}$,
and $\mathcal{S}_{m,k}^{(J)}= D_{m,k}^{(J)*} + D_{m,-k}^{(J)*}$. The quantum numbers $k$ and $m$
correspond to the projection, in units of $\hbar$, of the total angular momentum $J$ onto the
molecule-fixed $z$ axis and laboratory-fixed $Z$ axis, respectively. \eqref{eq:v_oc} retains only
the leading terms for Raman transitions restricted to $\Delta{k}=k-k'=0,\pm1$. The full expression
for the interaction potential including all off-diagonal polarizability terms can be found in the
supplementary material.

The handedness of the chiral field in \eqref{eq:v_ocdc} is defined by the relative signs of the dc
field $\epsilon_\text{dc}$ and the $\beta$ exponent, \ie, the direction of rotation of the optical
centrifuge. Replacing one enantiomer with another changes the $V^\text{oc}$ and $V^\text{dc}$
contributions in \eqref{eq:v_oc} and \eqref{eq:v_dc} as this changes the sign of one component of
the permanent molecular dipole moment as well as two components of the molecular polarizability
tensor~\cite{Yachmenev:PRL117:033001, Tutunnikov:JPCL9:1105}.

The simultaneous breaking of the rotational parity-inversion symmetry by the dc field and the
rotational axial symmetry by the optical centrifuge produces different rotational state populations
for the enantiomers. This enantiomer-specific effect can be explained using a minimal model of a
molecule as a three-level system of rotational states coupled by the dc and optical centrifuge
fields.
\begin{figure}
   \includegraphics[width=\linewidth]{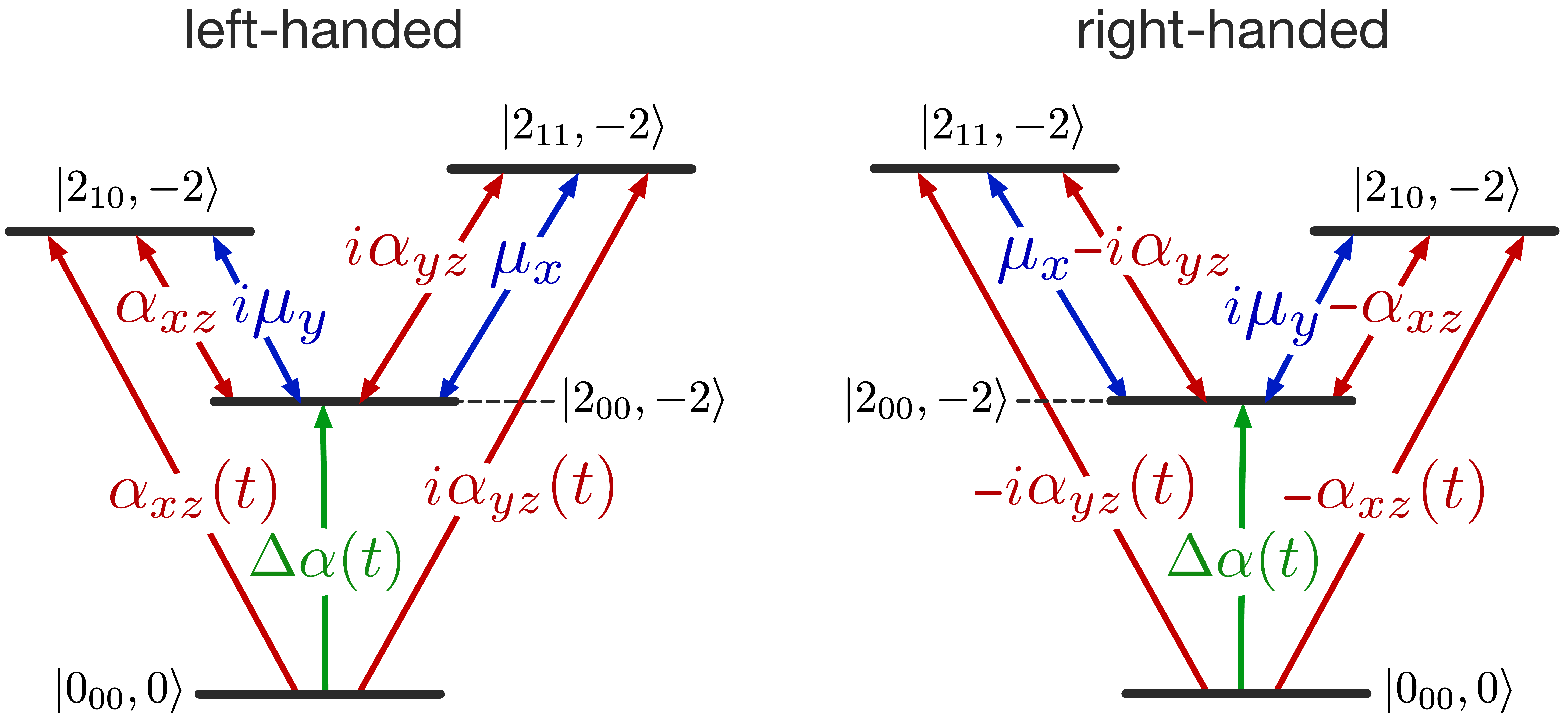}
   \caption{Minimal model of the enantiomer-differentiating transitions in a chiral molecule. A
      system of three-rotational states originating from the ground state is coupled to the static
      electric field and optical centrifuge, for instance, by the molecular electric dipole moment
      $\mu_y$ and the polarizability parameters $\Delta\alpha$ and $\alpha_{xz}$ or, alternatively,
      by $\mu_x$, $\Delta\alpha$, and $\alpha_{yz}$, see \eqref{eq:v_ocdc}--\eqref{eq:v_dc}. Here,
      $\Delta\alpha(t)=\Delta\alpha{}e^{2i\beta{}t^2}$ and
      $\alpha_{ab}(t)=\alpha_{ab}e^{2i\beta{}t^2}$, $a,b=x,y,z$.}
   \label{fig:model}
\end{figure}
An example of two such three-level systems is shown in \autoref{fig:model} for Raman transitions
from the ground state $\ket{J_{k\tau},m}=\ket{0_{00},0}$ to the excited states $\ket{2_{00},-2}$ and
$\ket{2_{10},-2}$, and from the ground to the $\ket{2_{00},-2}$ and $\ket{2_{11},-2}$ states. Note
that $\tau=0,1$ indicate the parity of the rotational wave function defined as $(-1)^\tau$. Assuming
that the different enantiomers are related by inversion along the molecular $z$ axis leads to a sign
change in the molecular dipole moment $\mu_z$ and the polarizabilities $\alpha_{xz}$ and
$\alpha_{yz}$ between the different enantiomers. For simplicity, different interaction terms in
\eqref{eq:v_oc} and \eqref{eq:v_dc} are abbreviated in \autoref{fig:model} with the corresponding
molecular polarizability $\Delta\alpha$, $\alpha_{xz}$, $\alpha_{yz}$ and dipole moment $\mu_x$,
$\mu_y$ symbols.

We consider one transition band formed by the ground state $|0_{00},0\rangle$ and the two excited
states $|2_{00},-2\rangle$ and $|2_{10},-2\rangle$ that, for simplicity, are assumed to have equal
energy $E$ relative to the ground state. When the angular acceleration of the centrifuge is
relatively slow, population transfer into both excited states, shown in \autoref{fig:model} with the
green and red colored upward arrows, will overlap in time: The optical centrifuge creates a
superposition of states
$\ket{\psi_0}=\cos(\chi{t})\ket{0_{00},0}+\Delta\alpha{t}e^{i\phi{t}}\ket{2_{00},-2}
+\alpha_{xz}{t}e^{i\phi{t}}\ket{2_{10},-2}$, with the phase angle
\mbox{$\chi=\epsilon_0^2\sqrt{\Delta\alpha^2+\alpha_{xz}^2}$} and
\mbox{$\phi=0.5(\alpha \epsilon_0^2-E)$}. The sign change of $\alpha_{xz}$ between enantiomers only
changes the phase of $|\psi_0\rangle$ by $\pi$, but does not alter any transition probabilities.
Note that we have also omitted the state superposition coefficients, which depend on the field
parameters.

The two excited states $|2_{00},-2\rangle$ and $|2_{10},-2\rangle$ are coupled to each other by
virtue of the induced dipole interaction with the rapidly oscillating field of the optical
centrifuge, $\alpha_{xz}\epsilon_0^2\equiv\mu_z^\text{ind}\epsilon_0$, and the permanent dipole
interaction with the dc field, $i\mu_y\epsilon_\text{dc}$. This quasi-static interaction, depicted
with the red and blue colored double-headed arrows in \autoref{fig:model}, creates the superposition
of states
$|\psi_0\rangle =\cos(\chi t) |0_{00},0\rangle + 0.5\alpha_{xz}t^2 c e^{i\theta} |2_{00},-2\rangle
+0.5\Delta\alpha t^2 c e^{-i\theta}|2_{10},-2\rangle$, with the coefficient
$c \sim |\mu_z^\text{ind}\epsilon_0+i\mu_y\epsilon_\text{dc}|$ and the phase angle
$\theta = \atantwo(\mu_y\epsilon_\text{dc}/\mu_z^\text{ind}\epsilon_0)$. Note that this expression
was obtained by truncating the exponential time-evolution operator at the second-order expansion
term and is only valid for short timescales.

The probability for the molecule to be, for example, in the $\ket{\mathit{2}}=\ket{2_{00},-2}$ state
is $P_\mathit{2}\propto\abs{\Delta\alpha{}e^{i\phi{}t}+0.5\alpha_{xz}t^2ce^{i\theta_{R/S}}}^2$ with
the sign of $\alpha_{xz}$ depending on the enantiomer and the phase angles $\theta_R$ and $\theta_S$
for the $R$ and $S$-enantiomers, respectively, which are related \emph{via}
$\theta_R=-\theta_S+\sign(\mu_y\epsilon_\text{dc})\pi$. The explicit expression for this probability
can be derived as
$P_\mathit{2}\propto\epsilon_0^4t^2(\Delta\alpha^2-\Delta\alpha\alpha_{xz}\mu_y\epsilon_\text{dc}t)+\mathcal{O}(t^4)$.
Similarly, the probability of being in the $\ket{\mathit{3}}=\ket{2_{10},-2}$ state is given by
$P_\mathit{3}\propto\epsilon_0^4t^2(\alpha_{xz}^2+\Delta\alpha\alpha_{xz}\mu_y\epsilon_\text{dc}t)+\mathcal{O}(t^4)$.
Here, the product of the mutually orthogonal polarizability and permanent dipole moment
$\alpha_{xz}\mu_y$ is independent of the choice of molecular axes, but changes sign between
enantiomers, which gives rise to different populations of the $R$ and $S$-enantiomers in the
$|2_{00},-2\rangle$ and $|2_{10},-2\rangle$ states.

Given the field configuration, \ie, the sign of the dc field $\pm\epsilon_\text{dc}$ and the
direction of centrifuge rotation $\pm\beta$, with $\pm\beta\equiv\pm{}t$ in the above equations, the
sign of the total product $\alpha_{xz}\mu_y\epsilon_\text{dc}t$ is determined by the absolute
structure of the molecule in the chosen reference system. The molecular electric dipole moment and
polarizability tensor can be obtained from an electronic structure calculation for a molecule of
interest. Thus, given the field configuration it is possible to predict the handedness of the
enantiomeric enhancement of the rotational state populations. The handedness can be reversed by
inverting the direction of either the dc field or the centrifuge rotation, see
\autoref{fig:concept}. Indeed, for the $R$-enantiomer of propylene oxide in the principal axes
system $\mu_y=0.69$, $\alpha_{xz}=-0.94$, and $\Delta\alpha=-6.26$, with all quantities in a.u.,
giving $P_\mathit{2}(R)<P_\mathit{2}(S)$ and $P_\mathit{3}(R)>P_\mathit{3}(S)$ for positive
$\epsilon_\text{dc}$ and $\beta$. This result is further confirmed by numerical simulations,
\emph{vide infra}.

From the expressions for $P_\mathit{2}$ and $P_\mathit{3}$, it is clear that a stronger dc field
increases the enantiomeric enhancement. However, it also induces strong Stark repulsion in the
excited states, which is neglected in our simple model, and hence decouple the two excited states in
the superposition $\ket{\psi_0}$ produced by the optical centrifuge. A simple condition has been
derived in the supplementary material to put an upper bound on the dc field for excited states of
interest. In principle, the use of a stronger dc field for higher rotational excitations will
improve the enantiomeric enhancement in the state populations, however, the difficulties associated
with spatially separating high-$J$ states with inhomogeneous electric fields must be taken into
consideration.

To quantitatively investigate the proposed scheme, full-dimensional quantum mechanical simulations
were performed for the prototypical chiral molecule propylene oxide (C$_3$H$_6$O). The computational
approach is based on highly accurate variational procedures capable of supporting high-resolution
spectroscopy, see \cite{Owens:JPCL9:4206} and \cite{Owens:PRL121:193201} for recent applications.
The field-free rotational motion was modeled using the rigid-rotor Hamiltonian with the rotational
constants $A=18023.89$~MHz, $B=6682.14$~MHz, and $C=5951.39$~MHz~\cite{Creswell:JMolSpec64:295,
   McGuire:Science352:1449}. The electric-dipole-moment vector and the static-polarizability tensor
were calculated \emph{ab initio} at the equilibrium geometry of the molecule using the coupled
cluster method CCSD(T) with the augmented correlation-consistent basis set
aug-cc-pVTZ~\cite{Dunning:JCP90:1007, Kendall:JCP96:6796} in the frozen-core approximation.
Electronic structure calculations employed the quantum chemistry package CFOUR~\cite{CFOUR:2017}.
Time-dependent quantum dynamics simulations used the computer program
RichMol~\cite{Owens:JCP148:124102}, which is a general-purpose code for modeling molecule-field
interactions. In simulations, the time-dependent wavefunction was built from a superposition of
field-free eigenstates and the time-dependent coefficients were obtained by numerical solution of
the time-dependent Schrödinger equation.

The optical centrifuge pulse was applied for a maximal duration of 140~ps with the peak amplitude of
the field $\epsilon_0=1\times10^7$~V/cm, chirp rate $\beta=(2\pi{}c)^2\cdot0.1~\text{cm}^{-2}$, and
carrier frequency of the field $\omega=c/(2\pi\cdot 800~\text{nm})$. The pulse envelope was modeled
using a half Gaussian profile with a $140$~ps half-width-at-half-maximum and a 1.5~ps smooth-cut
applied at the end of the truncated pulse. Several values for the dc field strength
$\epsilon_\text{dc}=1\ldots30$~kV/cm were investigated and optimal results were obtained for
$\epsilon_\text{dc}=5\ldots10$~kV/cm. Finite-initial-temperature effects were modeled by averaging
over the individual quantum wavepackets originating from different initial rotational states
according to Boltzmann statistics. Calculations were performed for rotational temperatures of
$T\leqslant1$~K, with $T\lessapprox1$~K directly achievable through supersonic expansion, whereas
effective temperatures $T\leqslant0.5$~K are representative of state-selected molecular
beams~\cite{Chang:IRPC34:557}.

\begin{figure}
   \includegraphics[width=\linewidth]{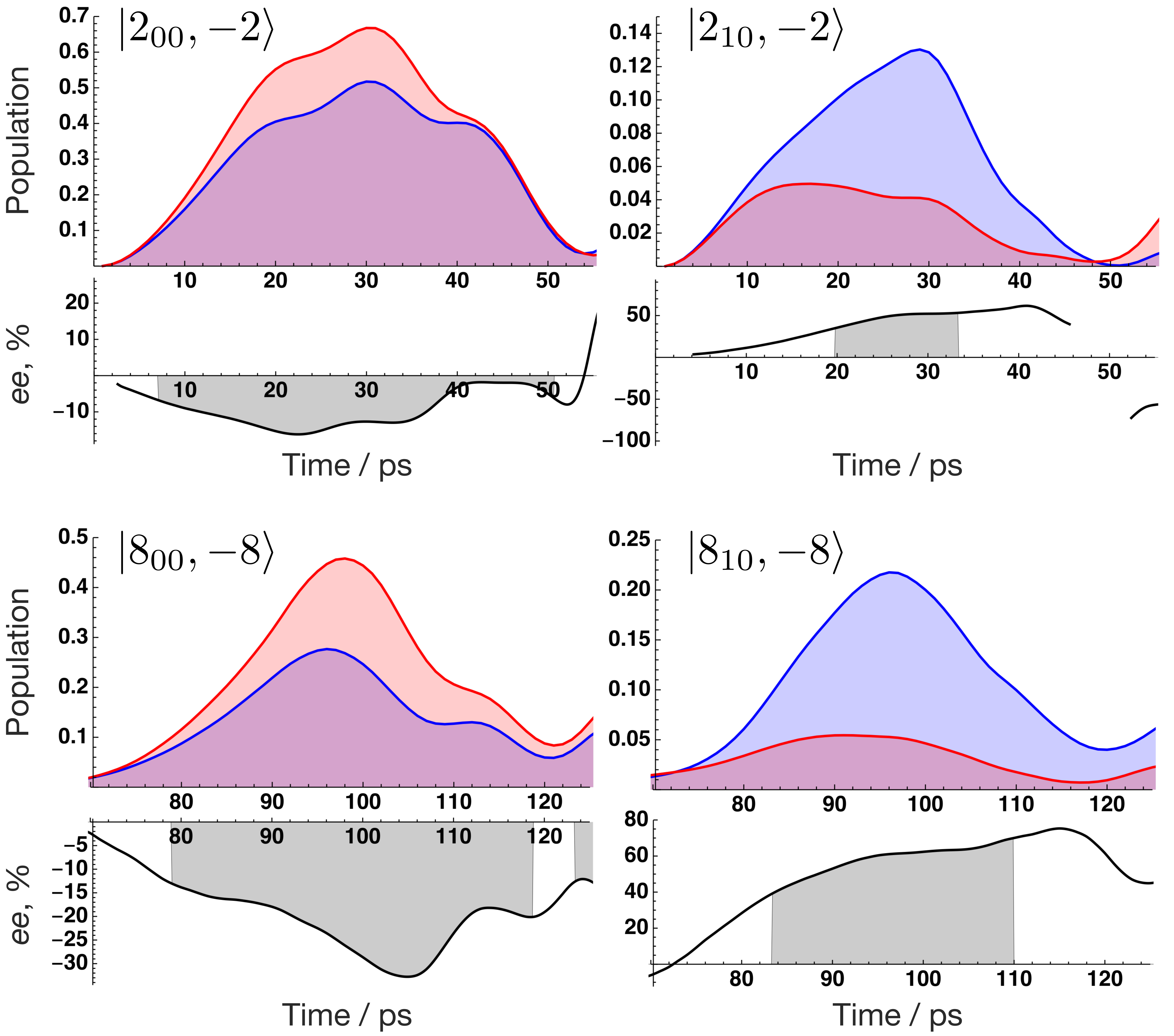}
   \caption{Temporal evolution of the rotational state populations for the lowest-energy excited
      state pair for $J=2$ and $J=8$ for the $R$ (blue) and $S$ (red) enantiomers of propylene
      oxide, for a dc field of 10~kV/cm. The enantiomeric excess $ee$ of the rotational-state
      populations is shown beneath each population plot. $ee$ is plotted only for times when the
      population of either of the enantiomers in the final state is greater than 1~\% of the initial
      population in $\ket{0_{00},0}$ at $t=0$. The shaded gray regions indicate corresponding
      populations greater than 10~\%.}
   \label{fig:temporal}
\end{figure}
The dynamics of the enantiomers in a chiral field with $\epsilon_\text{dc}=10$~kV/cm are depicted by
the time evolution of the rotational-state populations for the coupled pairs of the
$\ket{2_{00},-2}$ and $\ket{2_{10},-2}$ states and the $\ket{8_{00},-8}$ and $\ket{8_{10},-8}$
states, see \autoref{fig:temporal}. The initial rotational temperature was set to $T_i=0$~K, \ie,
all wavepacket population started in the rovibrational ground state \ket{0_{00},0}. The optical
centrifuge excited the molecules \emph{via} $\Delta{J}=2,\Delta{m}=-2$ rotational Raman transitions
and a significant amount of the wavepacket population for the $R$-enantiomer was transferred into
the $|2_{10},-2\rangle$ state at time $t\approx30$~ps and into the $|8_{10},-8\rangle$ state at
$t\approx100$~ps. This produced an excess of the $R$-enantiomer in the population of these states,
while an excess of the $S$-enantiomer is evident in the population of the $|2_{00},-2\rangle$ and
$\ket{8_{00},-8}$ states. To quantify the population difference, the time evolution of the
enantiomeric excess $\ee=(p(R)-p(S))/(p(R)+p(S))\cdot100~\%$ of the respective rotational state
populations $p(R)$ and $p(S)$ is shown in \autoref{fig:temporal}. As expected, the enantiomeric
enhancement is more pronounced at higher rotational excitations, owing to the smaller energy gap
between the $|8_{00},-8\rangle$ and $|8_{10},-8\rangle$ states. Furthermore, the absolute handedness
of the enantiomeric excess in these states confirms the predictions of the simplified three-level
model, \emph{vide supra}.

\begin{figure}
   \includegraphics[width=\linewidth]{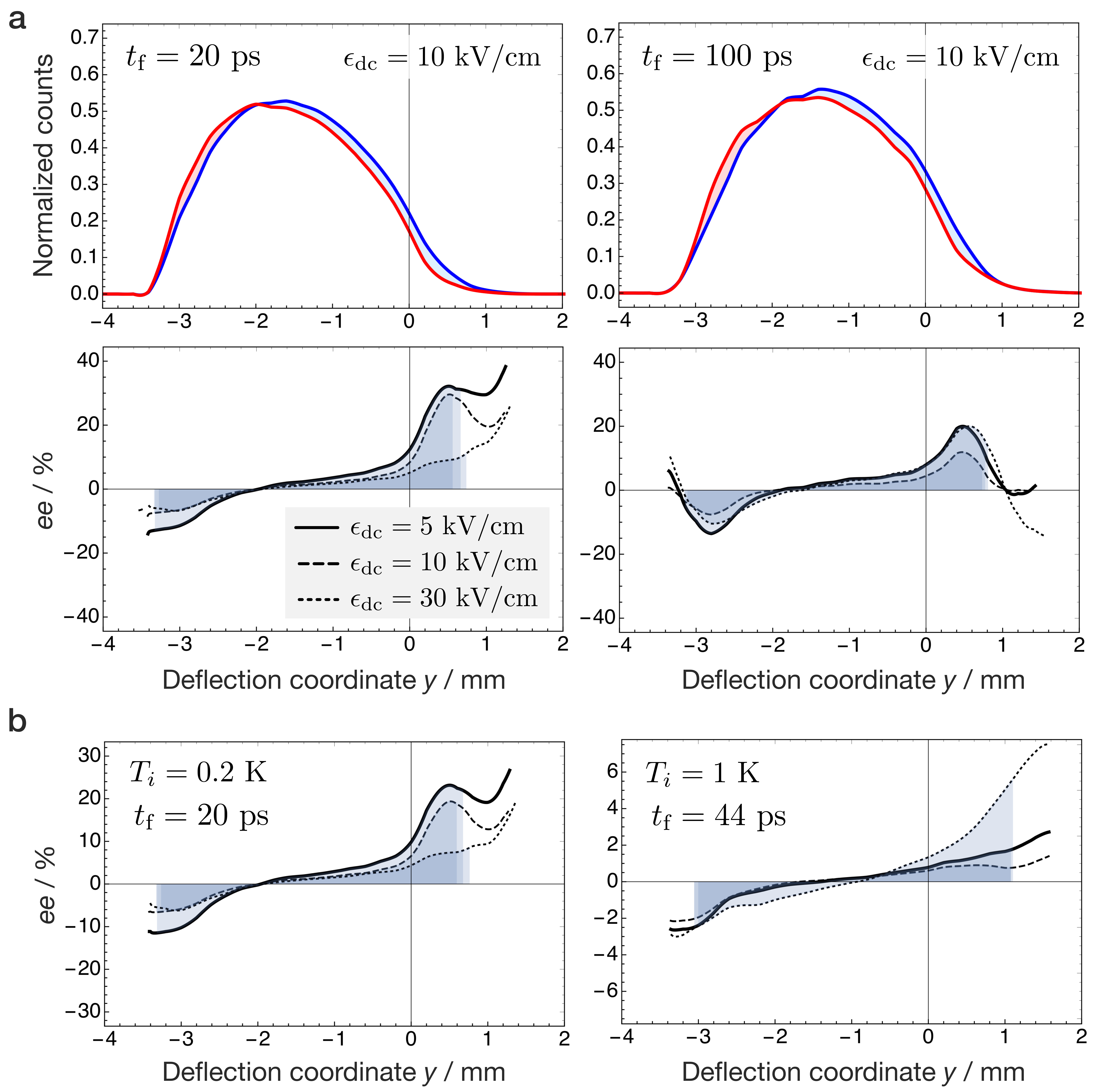}
   \caption{(a) Deflection profiles and enantiomeric excess \ee as a function of the vertical
      transverse beam position $y$, calculated for two release times from the optical centrifuge
      $t_\text{f}=20$~ps and $100$~ps, for the $R$ (blue) and $S$ (red) enantiomer of propylene
      oxide, for an initial rotational temperature of $T_i=0$~K. $y=0$ corresponds to the center of
      the undeflected molecular beam and the deflector's field strength increases in the negative
      $y$ direction, down to $y\approx-3$~mm~\cite{Kienitz:JCP147:024304}. (b) Enantiomeric excess
      \ee as a function of the deflection coordinate $y$, computed at release times
      $t_\text{f}=20$~ps and 44~ps for finite initial rotational temperatures of $T_i=0.2$~K and
      $1$~K, respectively. \ee is plotted only for deflection positions $y$ at which the ratio
      between the column density for either of the enantiomers and the total column density at
      $y=0$~mm is greater than 1~\%. Filled areas show ratios greater than 10~\%.}
   \label{fig:deflection}
\end{figure}
The enantiomeric enhancement in the rotational state populations can be converted to spatial
enrichment by releasing the molecules from the optical centrifuge at optimal times before entering
the electrostatic deflector, where molecules are spatially dispersed depending on their rotational
state. The rotational state populations of the $R$ and $S$-enantiomers were extracted from
quantum-mechanical simulations for different durations $t_\text{f}$ of the optical centrifuge pulse.
Then, their deflection profiles were simulated~\cite{Chang:CPC185:339, Filsinger:JCP131:064309} for
a typical experimental setup~\cite{Kienitz:JCP147:024304}, details are provided in the supplementary
material. The resulting vertical beam profiles are shown in \autoref{fig:deflection} for beams with
initial rotational temperatures of $T_i=0,0.2,1$~K.

In \autoref[a]{fig:deflection}, deflection profiles are shown for two optimal release times
$t_\text{f}=20$~ps and 100~ps for an initial $T_i=0$~K. Here, the difference in the profiles for the
$R$ and $S$-enantiomers are largely caused by the enhanced population of the $\ket{2_{00},-2}$ or
$\ket{8_{00},-8}$ state in the $S$-enantiomer, which both have stronger Stark shifts than the
respective $\ket{2_{10},-2}$ and $\ket{8_{10},-8}$ states, which produced an excess of the
$R$-enantiomer. Consequently, the $S$-enantiomer is more strongly deflected toward the larger dc
field, \ie, in the negative $y$ direction. Since population of the $\ket{2_{00},-2}$ or
$\ket{8_{00},-8}$ state in both enantiomers is larger than that of the $\ket{2_{10},-2}$ or
$\ket{8_{10},-8}$ state (see \autoref{fig:temporal}), deflection enhances the concentration of
$S$-enantiomers in the lower part of the deflected molecular beam, \ie, on the left-hand side of the
plot.

The spatial enantiomeric excess for different values of the dc field is illustrated for $T_i=0$~K
below the deflection profiles in \autoref[a]{fig:deflection}, and for $T_i=0.2$ and 1~K in
\autoref[b]{fig:deflection}. An optimal dc field between $\epsilon_\text{dc}=5$ and 10~kV/cm for
$T_i=0$ and 0.2~K provides maximal enantiomeric enrichment of about 30\%. The profile for
$T_i=0.2$~K is similar to that for $T_i=0$~K and is mostly dominated by the dynamics of the
$|2_{00},0\rangle$ and $|2_{10},0\rangle$ states with the absolute molecular intensity scaled by the
corresponding thermal Boltzmann factor. The enhancement diminishes for stronger dc fields of
$\epsilon_\text{dc}=30$~kV/cm due to the decoupling of the $|2_{00},0\rangle$ and $\ket{2_{10},0}$
states caused by strong Stark repulsion. For higher initial temperatures, excitation with the
optical centrifuge leads to a more spectrally broadened rotational wavepacket. For $T_i=1$~K the
wavepackets are composed of a large number of rotational states, with each state reaching its
largest $ee$ in population at a different time $t_\text{f}$. This significantly reduces the
enantiomeric excess to 6~\%. In addition, since many states with higher $k$ quanta are initially
populated, they exhibit smaller repulsive shifts and, therefore, display an increased $ee$ at
stronger dc fields. This is evidenced by the unexpected rise of enrichment at
$\epsilon_\text{dc}=30$~kV/cm for $T_i=1$~K.
The results of \autoref{fig:deflection} calculated for different release times $0<t_\text{f}<100$~ps
are shown through animated videos for different initial temperatures $T_i$ in the supplementary material.

In the presented scheme for the spatial separation of enantiomers, the enantiomeric enhancement is
associated with the distinct rotational excitation dynamics of the field-induced diastereomers of a
chiral molecule placed in a chiral field of defined handedness. The chiral field is created by the
combination of an optical centrifuge laser pulse with a dc field parallel or antiparallel to the
light's propagation direction. The enhancement of rotational state populations can be exploited to
spatially separate the $R$ and $S$ enantiomers using electrostatic deflection techniques. The
rotational-state-specific enantiomeric enhancement is fully controllable by changing the duration of
the optical centrifuge pulse. This has been confirmed through robust numerical simulations for
propylene oxide. A simple three-state model was established to explain these effects in a similar
manner to microwave three-wave mixing~\cite{Eibenberger:PhysRevLett118:123002, Perez:ACIE56:12512}.
In comparison to three-wave mixing, the presented approach offers much higher degrees of population
transfer, up to 80~\%, see \autoref{fig:temporal}, achieved by using a high-intensity chirped laser
field to drive the Raman-type transitions in a three-level system. Furthermore, our scheme can
clearly distinguish between right-handed and left-handed enantiomers and we envisage future
applications to the spatial separation of larger chiral molecules according to their absolute
configuration.

In principle, the presented scheme is applicable to any chiral polar molecule, but its
implementation would require optimization of the laser parameters and dc field
strength~\cite{Owens:JPCL9:4206, Kerbstad:NatComm10:658}. One of the more challenging considerations
is the optical centrifuge, which is commonly used to rotationally excite diatomic and triatomic
molecules up to extremely high angular momentum states, even up to
dissociation~\cite{Karczmarek:PRL82:3420, Villeneuve:PRL85:542, Hasbani:JCP116:10636}. Given that
low-$J$ state rotational excitation is sufficient, application to larger molecules should not be
overly problematic. However, large chiral molecules may require longer laser
pulses~\cite{Trippel:MP111:1738} to produce rotationally excited samples with sufficient densities.

This work has been supported by the Deutsche Forschungsgemeinschaft (DFG) through the priority
program ``Quantum Dynamics in Tailored Intense Fields'' (QUTIF, SPP1840, KU~1527/3, YA~610/1),
through the Clusters of Excellence ``Center for Ultrafast Imaging'' (CUI, EXC 1074, ID 194651731)
and ``Advanced Imaging of Matter'' (AIM, EXC 2056, ID 390715994), and by the European Research
Council under the European Union's Seventh Framework Programme (FP7/2007--2013) through the
Consolidator Grant COMOTION (ERC-614507-Küpper). J.O.\ and A.O.\ gratefully acknowledge fellowships
of the Alexander von Humboldt Foundation.

\bibliography{string,cmi}
\onecolumngrid
\end{document}